# Bridging Player Intentions: Exploring the Potential of Synchronized Haptic Controllers in Multiplayer Game

Kenta Hashiura[1,3], Kazuya Iida[2], Takeru Hashimoto[1], Youichi Kamiyama[3], Keita Watanabe[2], Kouta Minamizawa[3] and Takuji Narumi[1]

[1] *The University of Tokyo, Tokyo, Japan*
[2] *Meiji University, Tokyo, Japan*
[3] *Keio University, Tokyo, Japan*

(Email: k.hashiura@cyber.t.u-tokyo.ac.jp)

**Abstract ---** In multiplayer cooperative video games, players traditionally use individual controllers, inferring others' actions through on-screen visuals and their own movements. This indirect understanding limits truly collaborative gameplay. Research in Joint Action shows that when manipulating a single object, motor performance improves when two people operate together while sensing each other's movements. Building on this, we developed a controller allowing multiple players to operate simultaneously while sharing haptic sensations. We showcased our system at exhibitions, gathering feedback from over 150 participants on how shared sensory input affects their gaming experience. This approach could transform player interaction, enhance cooperation, and redefine multiplayer gaming experiences.

## 1 INTRODUCTION

The rapid advancement of information and communication technologies has led to a significant expansion of multiplayer gaming in the home console market. Multiplayer games have been shown to contribute positively to player enjoyment [1] and increase engagement with the game [2]. These games allow players to interact with others in virtual environments, creating shared experiences that extend beyond single-player gameplay. However, despite the social nature of these interactions, players often lack direct awareness of how their co-players are engaging with the game beyond what is visible on the shared screen.

Gaming consoles such as the Wii and Nintendo Switch offer multiplayer experiences where players share the same physical space and screen. These systems typically provide haptic feedback through their controllers, which has been demonstrated to enhance immersion in the gaming experience [3]. Nevertheless, this feedback is primarily designed to increase individual player immersion rather than facilitate awareness of other players' actions. As a result, players must rely primarily on visual cues from the shared screen to infer their co-players' engagement and strategies, potentially limiting the depth of their social gaming experience.

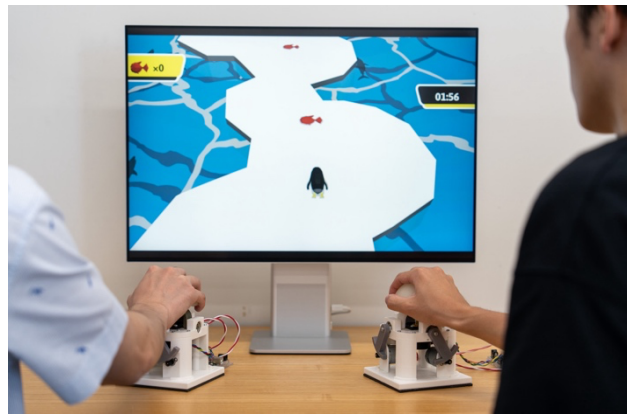

Fig.1 Playing ``Sliding Penguin Game``

In the field of joint action, where two people cooperate on a task, force and tactile feedback are known to improve performance. Ganesh et al. demonstrated that force feedback enhances performance in a task where two people control a cursor together [4]. Similarly, Jung et al. showed that tactile feedback improves the gaming experience for two players [5].

In this study, we developed a controller system that allows multiple players to share haptic feedback reflecting each other's joystick movements while playing games that require coordination of actions. We investigated how this synchronization of force feedback in video games influences player interactions, their ability to sense and respond to others' intentions, and overall gameplay experiences. Our research particularly focuses on exploring potential differences in these dynamics between two-player scenarios and those involving three or more players, aiming to contribute to a deeper understanding of how shared haptic information affects collaborative gameplay in multi-participant gaming environments.

## 2 Device

We have developed a force feedback controller that enables real-time sharing of operational sensations between multiple users. This system building upon the foundational work of Hashiura et al. [6] by integrating previously separate input and output devices into a single, cohesive unit. This system is not unique but has taken the orthodox form used in the development of joysticks with force feedback over the past two decades [7,8,9].

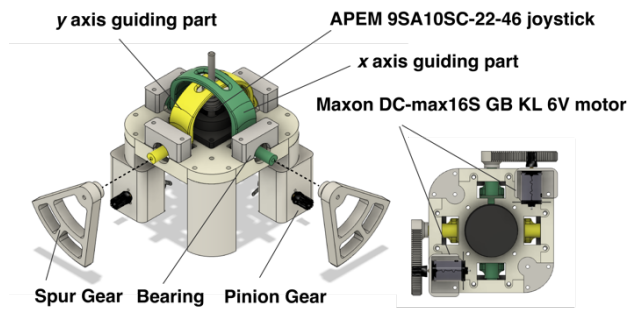

Fig.2 Device

The core of our system is a dual-axis force feedback joystick that allows free movement in both x and y directions. The controller's hardware comprises five key components: an APEM 9SA10SC-22-46 joystick for precise input, a Maxon DC-max16S GB KL 6V motor for force feedback, a Cytron 3Amp 4V-16V DC Motor Driver (2 Channels) for motor control, an ESP32-WROVER-E microcontroller for system management, and custom 3D-printed parts for structural integrity and enhanced haptic feedback.

The system's architecture follows a streamlined information flow, consisting of two main parts: input processing and force feedback output. For input processing, the ESP32-WROVER-E microcontroller continuously measures the joystick's angle. This angle data is then processed and transmitted to the game system.

In the force feedback output phase, the game system sends force feedback data based on the partner's input and game state. This data is received by the ESP32-WROVER-E, which converts it into appropriate current values. These current values are then sent to the motor driver to actuate the Maxon DC motor, providing force feedback to the user's joystick.

One of the unique features of our controller is its ability to create dynamic interactions between users. When operational sensations are shared, the partner's joystick movements are transmitted and reproduced through force feedback. This results in varying degrees of resistance depending on the users' actions. For instance, when users tilt their joysticks in the same direction, they experience minimal resistance, facilitating collaborative movements. Conversely, opposing movements generate strong resistance, providing a haptic sensation of conflict or opposition.

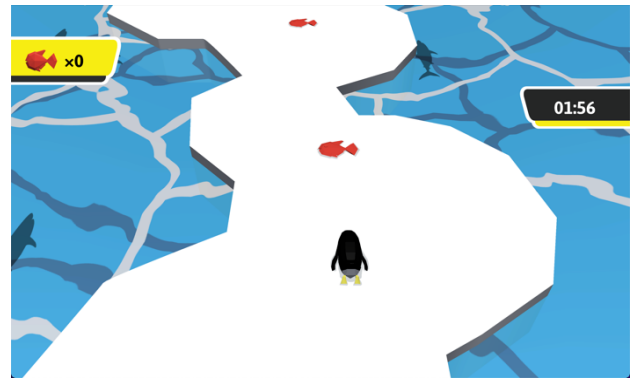

Fig.3 Game screen

## 3 Sliding Penguin Game

Sliding Penguin is one of the research games provided by the Open Video Game Library [10]. Players control a penguin sliding on ice, aiming to reach the goal without falling into the sea. The penguin is controlled using two-dimensional data representing the x and y directions. This data is used as acceleration, and the vector in the same direction increases as the penguin accelerates. Therefore, if you try to suddenly change direction while accelerating, you are more likely to lose control and fall into the sea due to its inertia. For this system, we modified the game to be controllable via OSC and designed it to be operated using multiple controllers.

## 4 Exhibition

We conducted two exhibitions, allowing over 150 participants to experience our system. The first exhibition featured two controllers, enabling participants to sense each other's movements. For the second exhibition, we expanded to four controllers, accommodating between two and four players simultaneously.

Participants first experienced the game without force feedback, followed by a session with haptic input sharing enabled. This setup allowed us to compare player experiences and decision-making processes under different conditions.

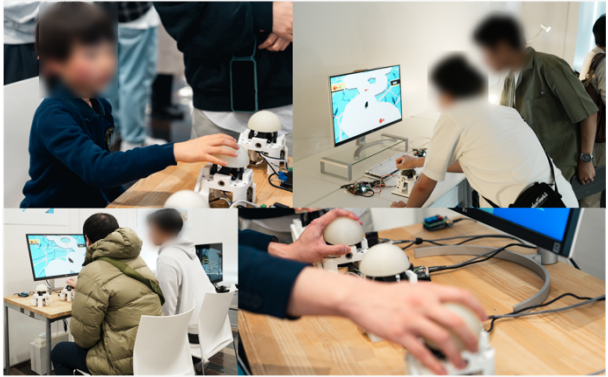

Fig.4 Exhibition

In two-player scenarios, participants noted that without haptic feedback, they occasionally struggled to control the penguin as intended. However, with haptic sharing, they could immediately sense conflicts in intentions. Some participants reported maintaining strong directional input to assert their desired path, while others adapted their movements to their partner's or assumed a "braking" role. These observations suggest that haptic feedback facilitated new decision-making processes not present in the non-haptic condition.

In three or four-player scenarios, participants expressed different sentiments. Some felt more confident in their actions when they could perceive others' movements through the controller. Others noted a sense of shared responsibility when input was synchronized, while some found it easier to choose their preferred direction without sharing.

This shift in perception between two-player and multi-player scenarios is intriguing. In pairs, players could maintain their input until their partner changed their decision. However, with three or more players, an individual's full input could be overridden by the combined input of many others.

Previous research has shown that multi-person control can lead to more stable movements [11], but the underlying decision-making processes remain unclear. Our exhibition results provide valuable insights into these dynamics, opening exciting avenues for future investigation into collaborative control and decision-making in shared-input gaming scenarios.

## 5 Demo at AsiaHaptics2024

For the upcoming demo at AsiaHaptics2024, we will prepare a gaming environment that accommodates up to four players simultaneously. The demo will be structured to highlight the impact of haptic feedback on collaborative gameplay and decision-making processes.

Participants will first experience the game without haptic feedback, allowing them to understand the challenges of interpreting other players' intentions solely through visual cues. Subsequently, they will play with haptic feedback enabled, experiencing how this additional sensory input affects their ability to collectively control the penguin's direction.

Through these experiences, we aim to demonstrate:
- In two-player scenarios: How players incorporate their partner's intentions into their own decision-making process.
- In three or more player scenarios: How individuals navigate between aligning with the group consensus and asserting their own preferences.

This demo will provide valuable insights into the role of haptic feedback in fostering collaboration and communication in multiplayer gaming environments, as well as its potential implications for group dynamics and decision-making processes.

## 6 Conclusion

We have developed a system that synchronizes haptic feedback in multi-participant gaming environments, allowing players to share operational sensations. Our exhibitions, involving over 150 participants, revealed that this shared sensory input affects gameplay experiences, with participants reporting an immediate sense of others' intentions and the emergence of new collaborative dynamics. Interestingly, our observations suggest potential differences in interaction patterns between two-player scenarios and those involving three or more players, with two-player settings showing more direct negotiations of control and multi-player scenarios introducing more complex group dynamics. These preliminary observations open up new questions about decision-making processes in collaborative gaming environments and contribute to our understanding of cooperative gameplay design. Our findings lay the groundwork for further investigation into how sensory sharing can enhance player engagement and interaction in multi-participant gaming experiences.

ACKNOWLEDGEMENT

This work was supported by JST SPRING, Grant Number JPMJSP2108.

REFERENCES

[1] Penelope Sweetser and Peta Wyeth. GameFlow: a model for evaluating player enjoyment in games. Comput. Entertain., Vol. 3, No. 3, pp. 3–3, July 2005.

[2] Alexander J Bisberg, Julie Jiang, Yilei Zeng, Emily Chen, and Emilio Ferrara. The gift that keeps on giving: Generosity is contagious in multiplayer on-line games. Proc. ACM Hum. Comput. Interact., Vol. 6, No. CSCW2, pp. 1–22, November 2022.

[3] Tanay Singhal and Oliver Schneider. Juicy haptic design: Vibrotactile embellishments can improve player experience in games. In Proceedings of the 2021 CHI Conference on Human Factors in Computing Systems, No. Article 126 in CHI '21, pp. 1–11, New York, NY, USA, May 2021. Association for Computing Machinery

[4] G Ganesh, A Takagi, R Osu, T Yoshioka, M Kawato, and E Burdet. Two is better than one: physical interactions improve motor performance in humans. Sci. Rep., Vol. 4, p. 3824, January 2014

[5] Sungchul Jung, Yuanjie Wu, Ryan McKee, and Robert W Lindeman. All shook up: The impact of floor vibration in symmetric and asymmetric immersive multi-user VR gaming experiences. In 2022 IEEE Conference on Virtual Reality and 3D User Interfaces (VR), pp. 737–745, March 2022.

[6] Kenta Hashiura, Youichi Kamiyama, Taku Tanichi, Mina Shibasaki, Keigo Inukai, and Minamizawa Kouta. Construction of tactile sharing device ui of the chicken game. Asia Haptics 2022, 2022.

[7] Byunghoon Bae, Taeoh Koo, Kyihwan Park, and Yongdae Kim. Design and control of a two degree of freedom haptic device for the application of PC video games. In Proceedings 2001 IEEE/RSJ International Conference on Intelligent Robots and Systems. Expanding the Societal Role of Robotics in the the Next Millennium (Cat. No.01CH37180), Vol. 3, pp. 1738–1743 vol.3. IEEE, 2002.

[8] Kyihwan Park, Byunghoon Bae, and Taeoh Koo. A haptic device for PC video game application. Mechatronics (Oxf.), Vol. 14, No. 2, pp. 227–235, March 2004.

[9] Youngbo Aram Shim, Keunwoo Park, Sangyoon Lee, Jeongmin Son, Taeyun Woo, and Geehyuk Lee. FSpad: Video game interactions using force feedback gamepad. In Proceedings of the 33rd Annual ACM Symposium on User Interface Software and Technology, UIST '20, pp. 938–950, New York, NY, USA, October 2020. Association for Computing Machinery.

[10] Kazuya Iida, Yuma Ina, Daichi Hayashi, Yohei Yanase, and Keita Watanabe. Open video game library: Developing a video game database for use in research and experimentation. In 29th ACM Symposium on Virtual Reality Software and Technology, New York, NY, USA, October 2023. ACM.

[11] Atsushi Takagi, Masaya Hirashima, Daichi Nozaki, and Etienne Burdet. Individuals physically interacting in a group rapidly coordinate their movement by estimating the collective goal. Elife, Vol. 8, , February 2019.